\documentclass[aps,
showpacs,pra]{revtex4}
\usepackage{amsmath,latexsym,amssymb}
\usepackage{amsmath}
\usepackage{latexsym}
\usepackage{amssymb}
\begin{document}
\title{Classification of local realistic theories}
\author{Koji Nagata}
\affiliation{
Department of Physics, Korea Advanced Institute of Science and Technology,
Daejeon 305-701, Korea
}
\date{\today}

\begin{abstract}

Recently, it has shown that an explicit local realistic model for the values 
of a correlation function, given in a two-setting Bell experiment 
(two-setting model), works only for the specific set of settings in the given experiment, but cannot construct a local realistic model for the values of a correlation function, given in a {\it continuous-infinite} settings 
Bell experiment (infinite-setting model), even though there exist two-setting models for all directions in space. 
Hence, two-setting model does not have the property which infinite-setting model has.
Here, we show that an explicit two-setting model 
cannot construct a local realistic model for 
the values of a correlation function, given in a {\it only discrete-three} settings Bell experiment (three-setting model), even though there exist two-setting models 
for the three measurement directions chosen in the given 
three-setting experiment. Hence, two-setting model does not have the property which three-setting model has.

\end{abstract}

\maketitle

\section{Introduction}

There is much research about local realism \cite{bib:Einstein, bib:Bell, bib:Redhead, bib:Peres3}. 
The locality condition says that a result of measurement pertaining to one system is independent of any measurement performed simultaneously at a distance on another system.
Quantum mechanics does not allow a local realistic interpretation. 
Certain quantum predictions violate Bell inequalities \cite{bib:Bell}, 
which are conditions that a local realistic theory must satisfy.
Experimental 
efforts (Bell experiment) of a violation of local realism 
can be seen in Refs.~\cite{experiment1,experiment2,experiment3}.
Other type of inequalities are given in Refs.~\cite{CH,W}.
Bell inequalities with settings other than spin polarizations can be seen in 
Ref.~\cite{Bramon}.

Here, we consider Bell experiment for a system described by
multipartite states in the case where
$n$-dichotomic observables are measured per site.
If $n$ is two, we consider two-setting Bell experiment.
If $n$ is three, we consider three-setting Bell experiment and so on.

Recently, it has shown \cite{Nagata1} that 
an explicit local realistic model for the values of 
a correlation function, given in a two-setting Bell experiment 
(two-setting model),
works only for the specific set of settings in the given experiment, but
cannot construct a local realistic model for the values of a correlation function, given in a Bell experiment with continuous-infinite settings lie in a 
plane (plane-infinite-setting model), even though there exist two-setting models for all directions in 
the plane. 
Therefore, the property of two-setting model is different from the property of
plane-infinite-setting model.

Further, in specific type of quantum states, it was shown \cite{Nagata2} that 
Bell inequality with the assumption of the existence of 
a local realistic model which is rotationally invariant 
(sphere-infinite-setting model)
disproves two-setting model stronger than 
Bell inequality with the assumption of the existence of 
a local realistic model
which is rotationally invariant with respect to a plane (i.e., plane-infinite-setting model).
Therefore, the property of two-setting model is different from the property of
sphere-infinite-setting model.
Also we see that the property of plane-infinite-setting model 
is different from the property of
sphere-infinite-setting model.

We thus see that there is a division among the measurement 
settings, those that admit local realistic models which are 
rotationally invariant (sphere-infinite-setting model),
those that admit local realistic models which are 
rotationally invariant with respect to a plane (plane-infinite-setting model),
and 
those that do not (e.g., two-setting model). 
This is another manifestation of the underlying 
contextual nature of local realistic theories of quantum experiments.

In this paper, we shall show that two-setting model cannot construct a 
local realistic model for the values of a correlation function, given in a three-setting Bell experiment (three-setting model), even though there exist two-setting models 
for the three measurement directions chosen in the given 
three-setting experiment. 
The property of two-setting model 
is different from the property of three-setting model.
To this end, we derive a generalized Bell inequality for $N$ qubits which involves three-setting for each of the local measuring apparatuses. The inequality forms a necessary condition for the existence of three-setting model.
Although the inequality involves three settings, 
it can be experimentally tested using two orthogonal local 
measurement settings.
This is a direct consequence of the assumed form of rotationally invariant correlation like (\ref{et}).
We see our generalized Bell inequality with the assumption of the existence of 
three-setting model
disproves two-setting model for the actually measured values of the correlation function.

Our result provides classification of local realistic theories.
In order to say that some model is different from another model, 
we need criterion.
Our criterion is as follows.
{\it We may say that model (A1) is different from model (A2) if model (A1) 
does not have the property which model (A2) has}.
We shall stand to this approach.

Then, we can see four types of models at least.
First, there is two-setting model.
It is explicitly constructed by standard two-setting Bell 
inequalities \cite{bib:Zukowski2}.
However, this model is disproved by several generalized Bell inequalities.
The patterns of the disqualification are different each other.
Therefore, one furthermore has three different types of models.
These are three-setting model, plane-infinite-model, 
and sphere-infinite-model, as we mentioned above.

\section{Multipartite three-setting generalized Bell inequality}

Assume that we have a set of $N$ spins $\frac{1}{2}$. Each of them is in a separate laboratory. As is well known the measurements (observables) for such spins are parameterized by a unit vector $\vec n_j,~j=1,2,\ldots,N$ (direction along which the spin component is  measured). The results of measurements are $\pm 1$ (in $\hbar/2$ unit). One can introduce the ``Bell'' correlation function, which is the average of the product of the local results:
\begin{equation}
E(\vec n_1, \vec n_2, ..., \vec n_N) = \langle r_1(\vec n_1) r_2(\vec n_2) ... r_N(\vec n_N) \rangle_{avg},
\end{equation}
where $r_j(\vec n_j)$ is the local result, $\pm 1$, which is obtained if the measurement direction is set at $\vec n_j$. 
If experimental correlation function admits a rotationally invariant tensor structure 
familiar from quantum mechanics, we can introduce the following form:
\begin{equation}
E(\vec n_1, \vec n_2, ..., \vec n_N) = \hat T \cdot (\vec n_1 \otimes \vec n_2 \otimes ... \otimes \vec n_N),
\label{et}
\end{equation}
where $\otimes$ denotes the tensor product, $\cdot$ the scalar product in $\mathrm{R}^{3N}$ and $\hat T$ is the correlation tensor given by
\begin{equation}
T_{i_1...i_N} \equiv E(\vec x_{1}^{(i_1)},\vec x_{2}^{(i_2)}, ..., \vec x_{N}^{(i_N)}),
\label{tensor}
\end{equation}
where $\vec x_{j}^{(i_j)}$ is a unit directional vector of the local coordinate system of the $j$th observer; $i_j = 1,2,3$ gives the full set of orthogonal vectors defining the local Cartesian coordinates. 
That is, the components of the correlation tensor are experimentally accessible by measuring the correlation function at the directions given by the basis vectors of local coordinate systems.
Obviously the assumed form of (\ref{et}) implies rotational invariance, because the correlation function is a scalar. Rotational invariance simply states that the value of $E(\vec n_1, \vec n_2, ..., \vec n_N)$ cannot depend on the local coordinate systems used by the $N$ observers. Assume that  one knows the values of all $3^N$ components of the correlation tensor, $T_{i_1...i_N}$, which are obtainable by performing specific $3^N$  measurements of the correlation function, compare Eq.~(\ref{tensor}). Then, with the use of formula (\ref{et}) one can reproduce the value of the correlation functions for all other possible sets of local settings.

Using this rotationally invariant structure of the correlation function, 
we shall derive a necessary condition for the existence of a local realistic 
model for the values of the experimental correlation function (\ref{et}) 
given in a three-setting Bell experiment.

If the correlation 
function is described by a local realistic theory,
then the correlation function must be simulated by the following structure
\begin{eqnarray}
E_{LR}(\vec{n}_1,\vec{n}_2,\ldots,\vec{n}_N)=
\int d\lambda \rho(\lambda)
I^{(1)}(\vec{n}_1,\lambda)I^{(2)}(\vec{n}_2,\lambda)\cdots
I^{(N)}(\vec{n}_N,\lambda),\label{LHVcofun}
\end{eqnarray}
where $\lambda$ is some local hidden variable, $\rho(\lambda)$ is a probabilistic distribution, and $I^{(j)}(\vec{n}_j,\lambda)$ is 
the predetermined ``hidden'' result of 
the measurement of the dichotomic observable 
$\vec n \cdot \vec \sigma$ with values $\pm 1$.

Let us parametrize the three unit vectors in the plane defined by $\vec x_j^{(1)}$ and $\vec x_j^{(2)}$ in the following way
\begin{eqnarray}
\vec{n}_j(\alpha_j^{l_j})=\cos \alpha_j^{l_j} \vec{x}_j^{(1)}
+\sin \alpha_j^{l_j} \vec{x}_j^{(2)}, ~j=1,2,\ldots,N.\label{vector}
\end{eqnarray}
The phases $\alpha_j^{l_j} $ that experimentalists are allowed to set are chosen as
\begin{eqnarray}
\alpha_j^{l_j}=(l_j-1)\pi/3,~ l_j=1,2,3.
\end{eqnarray}

We shall show that the scalar product 
of ``three-setting'' local realistic correlation function
\begin{eqnarray}
E_{LR}(\alpha_1^{l_1},\alpha_2^{l_2},\ldots,\alpha_N^{l_N})=
\int d\lambda \rho(\lambda)
I^{(1)}(\alpha_1^{l_1},\lambda)I^{(2)}(\alpha_2^{l_2},\lambda)\cdots
I^{(N)}(\alpha_N^{l_N},\lambda),\label{3model}
\end{eqnarray}
with rotationally invariant correlation function, that is
\begin{eqnarray}
E(\alpha_1^{l_1},\alpha_2^{l_2},\ldots,\alpha_N^{l_N})=
\hat{T} \cdot \vec{n}_1(\alpha_1^{l_1})
\otimes
\vec{n}_2(\alpha_2^{l_2})
\otimes\cdots
\otimes\vec{n}_N(\alpha_N^{l_N}),
\end{eqnarray}
is bounded by a specific number dependent on $\hat{T}$.
We define the scalar product $(E_{LR}, E)$ as follows. We see that the maximal 
possible 
value of $(E_{LR}, E)$ is bounded as:
\begin{eqnarray}
 (E_{LR}, E) =\sum_{l_1=1,2,3}
\sum_{l_2=1,2,3}\cdots
\sum_{l_N=1,2,3}
E_{LR}(\alpha_1^{l_1},\alpha_2^{l_2},\ldots,\alpha_N^{l_N})
E(\alpha_1^{l_1},\alpha_2^{l_2},\ldots,\alpha_N^{l_N})
\leq 2^N T_{max},\label{Bell-Zineq}
\end{eqnarray}
where $T_{max}$ is the maximal 
possible value of the correlation tensor component, i.e.,
\begin{eqnarray}
T_{max}\equiv\max_{\beta_1,\beta_2,\ldots,\beta_N}
E(\beta_1,\beta_2,\ldots,\beta_N),
\label{TE}
\end{eqnarray}
where $\beta_j$ is some angle.

A necessary condition 
for the existence of ``three-setting'' local realistic 
description $E_{LR}$ of the experimental correlation function 
\begin{equation}
E(\alpha_1^{l_1},\alpha_2^{l_2},\ldots,\alpha_N^{l_N})
=E(\vec n_1(\alpha_1^{l_1}),..., \vec n_N(\alpha_N^{l_N})),
\end{equation} 
that is for $E_{LR}$ to be equal to $E$ for the three measurement directions, 
is that one has $(E_{LR}, E) =(E, E)$.
This implies the possibility of modeling $E$ by ``three-setting'' local realistic correlation function $E_{LR}$ given in (\ref{3model}) with 
respect to the three measurement directions.
If we have $( E_{LR}, E)< (E, E)$,
then the experimental correlation 
function 
cannot be explainable by three-setting local realistic model. 
(Note that, due to the summation in (\ref{Bell-Zineq}), 
we are looking for three-setting model.)

In what follows, we derive the upper bound (\ref{Bell-Zineq}).
Since the local realistic model is an average over $\lambda$, it is enough to find the bound of the following expression
\begin{eqnarray}
\sum_{l_1=1,2,3}\cdots
\sum_{l_N=1,2,3}
I^{(1)}(\alpha_1^{l_1},\lambda)\cdots
I^{(N)}(\alpha_N^{l_N},\lambda) 
\times \sum_{i_1,i_2,\ldots,i_N=1,2}T_{i_1i_2...i_N}
c^{i_1}_{1}c^{i_2}_{2}\cdots c^{i_N}_{N},\label{integral}
\end{eqnarray}
where
\begin{eqnarray}
(c^1_{j}, c^2_{j})=(\cos \alpha_j^{l_j}, \sin \alpha_j^{l_j}),
\end{eqnarray}
and
\begin{eqnarray}
T_{i_1i_2...i_N}=\hat{T} \cdot
(\vec{x}_1^{(i_1)}\otimes\vec{x}_2^{(i_2)}\otimes
\cdots\otimes\vec{x}_N^{(i_N)}),
\end{eqnarray}
compare (\ref{et}) and (\ref{tensor}). 

Let us analyze the structure of this sum (\ref{integral}).
Notice that (\ref{integral}) is a sum, with coefficients given by
$T_{i_1i_2...i_N}$, which is a product of the following sums:
\begin{eqnarray}
\sum_{l_j=1,2,3} I^{(j)}(\alpha_j^{l_j}, \lambda) \cos \alpha_j^{l_j},
\end{eqnarray}
and
\begin{eqnarray}
\sum_{l_j=1,2,3} I^{(j)}(\alpha_j^{l_j}, \lambda) \sin \alpha_j^{l_j}.
\end{eqnarray}
We deal here with sums, or rather scalar products of
$I^{(j)}(\alpha_j^{l_j}, \lambda)$ with two orthogonal vectors.
One has
\begin{eqnarray}
\sum_{l_j=1,2,3} \cos \alpha_j^{l_j}\sin \alpha_j^{l_j}=0,
\end{eqnarray}
because,
\begin{eqnarray}
2 \times \sum_{l_j=1,2,3} \cos \alpha_j^{l_j}\sin \alpha_j^{l_j}
=\sum_{l_j=1,2,3} \sin 2\alpha_j^{l_j}=
{\rm Im}\left(\sum_{l_j=1,2,3} e^{i 2\alpha_j^{l_j}}\right).
\end{eqnarray}
Since $\sum_{l_j=1}^3 e^{i(l_j-1)(2/3)\pi}=0$, 
the last term vanishes.

Please note
\begin{eqnarray}
\sum_{l_j=1}^3 (\cos \alpha_j^{l_j})^2
=\sum_{l_j=1}^3 \frac{1+\cos 2\alpha_j^{l_j}}{2}=3/2
\end{eqnarray}
and
\begin{eqnarray}
\sum_{l_j=1}^3 (\sin \alpha_j^{l_j})^2
=\sum_{l_j=1}^3 \frac{1-\cos 2\alpha_j^{l_j}}{2}=3/2,
\end{eqnarray}
because,
\begin{eqnarray}
\sum_{l_j=1,2,3} \cos 2\alpha_j^{l_j}=
{\rm Re}\left(\sum_{l_j=1,2,3} e^{i 2\alpha_j^{l_j}}\right).
\end{eqnarray}
Since $\sum_{l_j=1}^3 e^{i(l_j-1)(2/3)\pi}=0$, 
the last term vanishes.

The normalized vectors 
$M_1 \equiv \sqrt{\frac{2}{3}}(\cos 0,\cos \pi/3,\cos 2\pi/3)$ and 
$M_2 \equiv \sqrt{\frac{2}{3}}(\sin 0,\sin \pi/3,\sin 2\pi/3)$ form a basis of a real two-dimensional plane, which we shall call $S^{(2)}$.
Note further that any vector in $S^{(2)}$ is of the form 
\begin{eqnarray}
A\cdot M_1 + B\cdot M_2,
\end{eqnarray}
where $A$ and $B$ are constants,
and that any normalized vector in $S^{(2)}$ is given by
\begin{eqnarray}
\cos \psi M_1 + 
\sin \psi M_2
=\sqrt{\frac{2}{3}}(\cos(0-\psi),
\cos (\pi/3-\psi),\cos (2\pi/3-\psi)).
\end{eqnarray}
The norm 
$\Vert I^{(j){||}} \Vert$ of the projection 
of $I^{(j)}$ into the 
plane $S^{(2)}$ is given by the maximal possible 
value of the scalar product 
$I^{(j)}$ with any normalized vector belonging to $S^{(2)}$, that is
\begin{eqnarray}
\Vert I^{(j){||}} \Vert= \max_{\psi}
\sum_{l_j=1,2,3}
I^{(j)}(\alpha_j^{l_j}, \lambda)
\sqrt{\frac{2}{3}}\cos (\alpha_j^{l_j}-\psi)
=\sqrt{\frac{2}{3}}\max_{\psi}{\rm Re}
\left(z\exp\left(i(-\psi)\right)\right).
\end{eqnarray}
where $z=\sum_{l_j=1}^3 I^{(j)}(\alpha_j^{l_j}, \lambda)
\exp\left(i \alpha_j^{l_j}\right)$.
We may 
assume $|I^{(j)}(\alpha_j^{l_j}, \lambda)|=1$. 
Then, since $e^{i\alpha_j^{l_j}}= e^{i[(l_j-1)/3]\pi}$, the
possible values for $z$ are $0, \pm 2e^{i(\pi/3)}, \pm
2e^{i(2\pi/3)}, \pm 2$. Note that 
the minimum possible overall complex phase (modulo $2\pi$) of
$\left(z\exp\left(i(-\psi)\right)\right)$ is $0$. Then we obtain
$
\Vert I^{(j){||}} \Vert\leq \sqrt{\frac{2}{3}}\times 2 \cos 0
=2\sqrt{\frac{2}{3}}.$
That is, one has $\Vert I^{(j){||}} \Vert\leq 2\sqrt{\frac{2}{3}}$.

Since $M_1$ 
and $M_2$ are two orthogonal 
basis vectors in $S^{(2)}$, one has
\begin{eqnarray}
\sum_{l_j=1,2,3} 
I^{(j)}(\alpha_j^{l_j}, \lambda)\cdot
\sqrt{\frac{2}{3}}\cos \alpha_j^{l_j}=\cos \beta_j \Vert I^{(j){||}} \Vert,
\end{eqnarray}
and
\begin{eqnarray}
\sum_{l_j=1,2,3}  
I^{(j)}(\alpha_j^{l_j}, \lambda)\cdot
\sqrt{\frac{2}{3}}\sin \alpha_j^{l_j}=\sin \beta_j \Vert I^{(j){||}} \Vert,
\end{eqnarray}
where $\beta_j$ is some angle.
Using this fact one can put the value of (\ref{integral}) into the following form
\begin{eqnarray}
\left(\sqrt{\frac{3}{2}}\right)^{N}\prod_{j=1}^N \Vert I^{(j){||}} \Vert
\times \sum_{i_1,i_2,\ldots,i_N=1,2}
T_{i_1i_2...i_N}
d^{i_1}_{1}d^{i_2}_{2}\cdots d^{i_N}_{N},
\label{EE}
\end{eqnarray}
where
\begin{eqnarray}
(d^{1}_j, d^{2}_j)=(\cos \beta_j, \sin \beta_j).
\label{dcos}
\end{eqnarray}

Let us look at the expression 
\begin{equation}
\sum_{i_1i_2...i_N=1,2}
T_{i_1i_2...i_N}
d^{i_1}_{1}d^{i_2}_{2}\cdots d^{i_N}_{N}.
\label{onT}
\end{equation}
Formula (\ref{dcos}) shows that we deal here with two dimensional unit vectors $\vec d_j = (d_j^1, d_j^2), j=1,2,...,N$, that is (\ref{onT}) is equal to $\hat T \cdot (\vec d_1 \otimes \vec d_2 \otimes .... \otimes \vec d_N)$, i.e., it is a component of the tensor $\hat T$ in the directions specified by the vectors $\vec d_j$. If one knows all the values of $T_{i_1i_2...i_N}$, one can always find the maximal possible value of such a component, and it is equal to $T_{max}$, of Eq.~(\ref{TE}).

Therefore since $\Vert I^{(j){||}} \Vert\leq 2\sqrt{\frac{2}{3}}$
the maximal value of (\ref{EE}) is $2^N T_{max}$, 
and finally one has
\begin{eqnarray}
(E_{LR}, E) \leq 2^N T_{max}.\label{fizu}
\end{eqnarray}

Please note that the relation (\ref{fizu}) is a generalized Bell inequality. 
Specific local realistic models, which predict three-setting models, must satisfy it. In the next section, we shall show that if one replaces $E_{LR}$ by $E$ one may have a violation of the inequality (\ref{fizu}). One has
\begin{eqnarray}
(E, E)=
\sum_{l_1=1,2,3}  
\sum_{l_2=1,2,3}\cdots
\sum_{l_N=1,2,3} 
\left(\sum_{i_1,i_2,\ldots,i_N=1,2}T_{i_1i_2...i_N}
c^{i_1}_{1}c^{i_2}_{2}\cdots c^{i_N}_{N}\right)^2 
=\left(\frac{3}{2}\right)^N \sum_{i_1,i_2,\ldots,i_N=1,2}T_{i_1i_2...i_N}^2.
\label{EEvalue}
\end{eqnarray}
Here, we have used the fact that $\sum_{l_j=1,2,3} ~ c_j^{i_1} c_j^{i_1'} 
= \frac{3}{2} \delta_{i_1i_1'}$, because 
$c_j^1 = \cos \alpha_j^{l_j}$ and $c_j^2 = \sin{\alpha_j^{l_j}}$.

The structure of condition (\ref{fizu}) and the value (\ref{EEvalue}) suggests that the value of (\ref{EEvalue}) does not have to be smaller than (\ref{fizu}). That is there may be such correlation functions $E$, which have the property that for any $E_{LR}$ (three-setting model) one has $(E_{LR}, E) < (E,E)$, which implies impossibility of modeling $E$ by ``three-setting'' local realistic correlation function $E_{LR}$ with respect to the three measurement directions.

\section{Difference between two-setting model and three-setting model}

We present here an important example of a violation of (\ref{fizu}). 
It presents the difference between two-setting model and three-setting model.
Imagine $N$ observers who can choose between two orthogonal directions of spin measurement, $\vec x_j^{(1)}$ and $\vec x_j^{(2)}$ for the $j$th one. Let us assume that the source of $N$ entangled spin-carrying particles emits them in a state, which can be described as a generalized Werner state, namely $V |\psi_{GHZ} \rangle \langle
\psi_{GHZ}|+(1-V) \rho_{noise}$, where $|\psi_{GHZ}
\rangle = 1/\sqrt{2} (|+\rangle_1\cdots|+\rangle_N +
|-\rangle_1 \cdots|-\rangle_N)$ is the 
Greenberger, Horne, and Zeilinger (GHZ) state \cite{bib:GHZ} and
$\rho_{noise} = \frac{1}{2^N} \openone$ is the random noise admixture. The value of $V$ can be interpreted as the reduction factor of the interferometric contrast observed in the multi-particle correlation experiment. The states $| \pm \rangle_j$ are the eigenstates of the $\sigma_z^j$ observable. One can easily show that if the observers limit their settings to $\vec x_j^{(1)} = \hat x_j$ and $\vec x_j^{(2)} = \hat y_j$ there are $2^{N-1}$ components of $\hat T$ of the value $\pm V$. These are $T_{11...1}$ and all components that except from indices 1 have an even number of indices 2. Other x-y components vanish.

It is easy to see that $T_{max}=V$ and 
$\sum_{i_1,i_2,\ldots,i_N=1,2}T_{i_1i_2...i_N}^2=V^2 2^{N-1}$.
Then, we have $(E_{LR}, E) \leq 2^N V$ 
and $(E,E) = (\frac{3}{2})^N V^2 2^{N-1}=\frac{3^N}{2}V^2$.
For $N \geq 6$, and $V$ given by
\begin{equation}
2\left(\frac{2}{3}\right)^N<V \leq \frac{1}{\sqrt{2^{N-1}}}
\end{equation}
despite the fact that there exist ``two-setting'' local realistic models for
three measurement directions in consideration 
($(0,\frac{\pi}{3},\frac{2\pi}{3})\equiv (A,B,C)$), these models cannot construct 
``three-setting'' local realistic models. 
Namely, even though there exist two-setting models for a set of measurement 
directions $(A,B)$, $(B,C)$, and $(C,A)$, these models cannot construct 
any three-setting models for $(A,B,C)$.

As it was shown in \cite{bib:Zukowski2} if the correlation tensor satisfies the following condition
\begin{equation}
\sum_{i_1,i_2,...,i_N=1,2} T_{i_1i_2...i_N}^2 \leq 1
\label{ZB}
\end{equation}
then there always exists ``two-setting'' local realistic model for the set of correlation function values for all directions lie in a plane. For our example the condition (\ref{ZB}) is met whenever $V \leq \frac{1}{\sqrt{2^{N-1}}}$. Nevertheless such models 
cannot construct ``three-setting'' local realistic models for $V > 
2\left(\frac{2}{3}\right)^N$. 
Thus the situation is such for 
$V \leq \frac{1}{\sqrt{2^{N-1}}}$ for all two settings per observer
experiments  one can construct ``two-setting'' local realistic
model for the values of the correlation function for the settings chosen in the experiment. 
One wants to construct ``three-setting'' local realistic model for three measurement directions $(A,B,C)$ using 
``two-setting'' local realistic models, $(A,B), (B,C),$ and $(C,A)$.
But these three ``two-setting'' models must be consistent with each other,
if we want to construct truly ``three-setting'' local realistic models
 beyond the $2^N$ settings to which each
of them pertains.
Our result clearly indicates that this is impossible for $V > 
2\left(\frac{2}{3}\right)^N$.
These ``two-setting'' local realistic models, $(A,B), (B,C),$ and $(C,A)$ must
contradict each other. Rather they are therefore invalidated.
In other words the explicit two-setting models, given in \cite{bib:Zukowski2}
work only for the specific set of settings in the given experiment, but
cannot construct a local realistic model for 
the values of a correlation function, given in a three-setting Bell experiment (three-setting model), even though there exist two-setting models 
for the three measurement directions chosen in the given 
three-setting experiment.

One can see that three-setting model (even if exists) 
does not have the property which plane-infinite-model has
when $2\left(\frac{2}{3}\right)^N>V> 
2\left(\frac{2}{\pi}\right)^N$ \cite{Nagata1}.
Thus, three-setting model is different from plane-infinite-model.

Please note that all information needed to get this conclusion can be obtained in a two-orthogonal-setting-per-observer experiments, that is with the information needed in the case of ``standard'' two settings Bell inequalities \cite{bib:Zukowski2,bib:Mermin,Belinskii,bib:Werner3}. To get both the value of (\ref{EEvalue}) and of $T_{max}$ it is enough to measure all values of $T_{i_1i_2...i_N}$, $i_1, i_2, ..., i_N=1,2$. 

\section{Summary}

In summary we derived a generalized Bell inequality for $N$ qubits which involves three-setting for each of the local measuring apparatuses. The inequality forms a necessary condition for the existence of a local realistic model for the values of a correlation function, given in a three-setting Bell experiment.
And we have shown that a local realistic model for the values of a correlation function,
 given in a two-setting Bell experiment, cannot construct a 
local realistic model for the values of a correlation function, given in a three-setting Bell experiment, even though there exist two-setting models 
for the three measurement directions chosen in the given 
three-setting experiment. 
Hence the property of two-setting model 
is different from the property of three-setting model.

Our result provided classification of local realistic theories.
At least, we can see four types of models.
First, there is two-setting model.
It is explicitly constructed.
However, this model is disproved by several generalized Bell inequalities.
The patterns of the disqualification are different each other.
Therefore, one furthermore has three different types of models.
These are three-setting model, plane-infinite-model, 
and sphere-infinite-model.

How does our Bell inequality help us to understand certain quantum protocols?
What can it be used for?
We leave these questions as an open question.

\acknowledgments

This work has been supported by Frontier Basic Research Programs at
KAIST and K.N. is supported by the BK21 research professorship.


\begin{thebibliography}{9}
\bibitem{bib:Einstein}
A. Einstein, B. Podolsky, and N. Rosen, Phys. Rev. {\bf 47}, 777 (1935).
\bibitem{bib:Bell}
J. S. Bell, Physics {\bf 1}, 195 (1964).
\bibitem{bib:Redhead}
M. Redhead,
{\it Incompleteness, Nonlocality, and Realism}, Second Impression, 
(Clarendon Press, Oxford, 1989).
\bibitem{bib:Peres3}
A. Peres, 
{\it Quantum Theory: Concepts and Methods}
(Kluwer Academic, Dordrecht, The Netherlands, 1993).

\bibitem{experiment1}
A. Aspect, P. Grangier, and G. Roger,
Phys. Rev. Lett. {\bf 47}, 460 (1981).
\bibitem{experiment2}
A. Aspect, P. Grangier, and G. Roger,
Phys. Rev. Lett. {\bf 49}, 91 (1982).
\bibitem{experiment3}
A. Aspect, J. Dalibard, and G. Roger,
Phys. Rev. Lett. {\bf 49}, 1804 (1982).

\bibitem{CH}
J. F. Clauser and M. A. Horne,
Phys. Rev. D {\bf 10}, 526 (1974).

\bibitem{W}
E. P. Wigner,
Am. J. Phys. {\bf 38}, 1005 (1970).

\bibitem{Bramon}
A. Bramon and M. Nowakowski,
Phys. Rev. Lett. {\bf 83}, 1 (1999).

\bibitem{Nagata1}
K. Nagata, W. Laskowski, M. Wie{\'s}niak, and M. \.Zukowski, 
Phys. Rev. Lett. {\bf 93}, 230403 (2004).

\bibitem{Nagata2}
K. Nagata, 
J. Phys. A: Math. Theor. {\bf 40}, 13101 (2007).

\bibitem{bib:Zukowski2}
M. \.Zukowski and \v{C}. Brukner, 
Phys. Rev. Lett. {\bf 88}, 210401 (2002).

\bibitem{bib:GHZ}
D. M. Greenberger, M. A. Horne, and A. Zeilinger,
in {\it Bell's Theorem, Quantum Theory and Conceptions of the Universe},
edited by M. Kafatos (Kluwer Academic, Dordrecht, The Netherlands, 
1989), pp. 69-72.

\bibitem{bib:Mermin}
N. D. Mermin, Phys. Rev. Lett. {\bf 65}, 1838 (1990).
\bibitem{Belinskii}
A. V. Belinskii and D. N. Klyshko,
Phys. Usp. {\bf 36}, 653 (1993).
\bibitem{bib:Werner3}
R. F. Werner and M. M. Wolf, 
Phys. Rev. A {\bf 64}, 032112 (2001).






\end{thebibliography}
\end{document}